\documentclass[draft,
	twocolumn,
preprintnumbers,
amsmath,
	amssymb,
	aps,
pre,
floatfix,
	fleqn,
]{revtex4-1}

  \usepackage{dynlearn}

  \graphicspath{{./img/}}
  \usepackage{physics} \usepackage{multirow}
  \usepackage{cleveref}
\usepackage{enumitem}
      \usepackage{soul}
      \usepackage{mathtools}
      \usepackage{cancel}
      \usepackage{graphicx}
      \usepackage{xcolor}
      \usepackage{changepage}
      \usepackage{textcomp}
      \usepackage{nccmath}
      \usepackage{amsthm}

\theoremstyle{definition}

\newcommand{\defn}{\stackrel{.}{=}} \newcommand{\kb}{k_\textnormal{B}} \newcommand{\ex}{\mathrm{e}} \newcommand{\param}{\lambda} \newcommand{\xx}{z}         \newcommand{\Wex}{W_\textnormal{ex}} \newcommand{\Qex}{Q_\textnormal{ex}} \newcommand{\Qhk}{Q_\textnormal{hk}} \newcommand{\gWex}{\mathcal{W}_\textnormal{ex}} \newcommand{\gQex}{\mathcal{Q}_\textnormal{ex}} \newcommand{\gQhk}{\mathcal{Q}_\textnormal{hk}} \newcommand{\Feq}{F^\textnormal{eq}} \newcommand{\Fnss}{\mathcal{F}^\textnormal{nss}}
      
      \newcommand{\Mgen}{\mathbf{M}^{\left( \textnormal{out} \middle| \textnormal{in}\right)}} 

\newcommand{\pr}[1]{\mathrm{Pr}\left(#1\right)} \newcommand{\cpr}[2]{\mathrm{Pr}\left(#1 \mid #2\right)} \newcommand{\Hent}[1]{H\bqty{#1}}  \renewcommand{\DKL}[2]{D_\textnormal{KL}\bqty{#1 \, \| \, #2}}  \newcommand{\Imut}[2]{I\bqty{#1 : #2}}

\begin{document}

\def\ourTitle{First and Second Laws of Information Processing\\
by Nonequilibrium Dynamical States}

\def\ourAbstract{The averaged steady-state surprisal links a driven stochastic system's
information processing to its nonequilibrium thermodynamic response. By
explicitly accounting for the effects of nonequilibrium steady states, a
decomposition of the surprisal results in an information processing First Law
that extends and tightens---to strict equalities---various information
processing Second Laws. Applying stochastic thermodynamics' integral fluctuation
theorems then shows that the decomposition reduces to the Second Laws under
appropriate limits. In unifying them, the First Law paves the way to identifying
the mechanisms by which nonequilibrium steady-state systems leverage
information-bearing degrees of freedom to extract heat. To illustrate, we
analyze an autonomous Maxwellian information ratchet that tunably violates
detailed balance in its effective dynamics. This demonstrates how the presence
of nonequilibrium steady states qualitatively alters an information engine's
allowed functionality.
}

\def\ourKeywords{rate equations, stochastic process, Markov model, information processing,
Markov chain, entropy production, reversibility
}

\hypersetup{
  pdfauthor={Mikhael Semaan},
  pdftitle={\ourTitle},
  pdfsubject={\ourAbstract},
  pdfkeywords={\ourKeywords},
  pdfproducer={},
  pdfcreator={}
}

\author{Mikhael T. Semaan}
\email{msemaan@ucdavis.edu}
\affiliation{Complexity Sciences Center and Department of Physics and Astronomy, University of California, Davis, One Shields Avenue, Davis, CA 95616}

\author{James P. Crutchfield}
\email{chaos@ucdavis.edu}
\affiliation{Complexity Sciences Center and Department of Physics and Astronomy, University of California, Davis, One Shields Avenue, Davis, CA 95616}

\bibliographystyle{unsrt}

\title{\ourTitle}

\begin{abstract}
  \ourAbstract
\end{abstract}

\keywords{\ourKeywords}

\preprint{\arxiv{2211.05849}}

\title{\ourTitle}
\date{\today}
\maketitle

\setstretch{1.0}

\section{Introduction}
\label{sec:intro}

In 1861, Maxwell introduced a thought experiment in which a ``very
neat-fingered being'' leveraged  observations to control a system that
violated the Second Law of thermodynamics \cite{maxwellTheoryHeat1888}. A
century later, attempting to resolve the paradox, Landauer quantitatively
bounded the requisite thermodynamic resources for erasing a single bit of
information in a physical information-bearing degree of freedom \cite{Land61a}.
These results have since stimulated many explorations of the fundamental
physics tying a system's thermodynamic behavior to its functioning as
an information processor \cite{Leff02a}.

One particular line of inquiry focused on \emph{autonomous Maxwellian
ratchets}. In this, a ratchet embedded in a thermal environment moves along an
\emph{information tape}, interacting with a single tape symbol at a time. The
information in the tape's cells  modifies the ratchet's statistical properties
while the ratchet absorbs and dissipates energy
\cite{mandalWorkInformationProcessing2012}. Recent results introduced an
\emph{information processing Second Law} (IPSL) for such systems that bounds
the asymptotic rate $\dot{\expval{W}}$ of extracted work
\cite{boydIdentifyingFunctionalThermodynamics2016}:
\begin{align}
  \beta \dot{\expval{W}} \leq \Delta h_\mu
  \textnormal{,}
  \label{eq:alec_IPSL}
\end{align}
where $\Delta h_\mu = h'_\mu - h_\mu$, $h'_\mu$ is the Shannon entropy rate of
the statistical process generating the output tape, $h_\mu$ is the same for the
input tape, and $\beta$ is the inverse temperature of the thermal environment.

Notably, the IPSL bound corrected previous ``single-symbol'' relations by
accounting explicitly for arbitrary-order temporal correlations in the input
and output symbol strings. This, then, led to the discovery that removing such
correlation increases the system's capacity to produce work---despite the
ratchet interacting with only a single symbol at a time.

More recently, Ref.~\cite{heInformationProcessingSecond2022} developed a similar
IPSL not for the asymptotic rate of extracted work from an infinite tape, but
for the finite-time ensemble-averaged work extracted when operating on a finite
tape:
\begin{align}
  \beta \expval{W} \leq \Delta \Hent{Z}
  \textnormal{,}
  \label{eq:intro_IPSL_finite}
\end{align}
where $Z$ is the random variable associated with the joint space of the ratchet
and tape and $\Hent{Z}$ is its Shannon entropy.

The following first derives a simple information-thermodynamic equality by
considering the averaged steady-state surprisal of a general driven stochastic
process:
\begin{align}
  \Delta \Hent{Z} = \expval{\gWex} - \expval{\gQex} - \Delta \DKL{Z}{\Lambda}
  \textnormal{.}
  \label{eq:IPFL_intro}
\end{align}
Here, $Z$ is the random variable associated with a system's state, $\Lambda$
that associated with an environmentally-induced steady state, and
$\expval{\gWex}$ and $\expval{\gQex}$ are (in entropic units) the average excess
work and heat of nonequilibrium steady state thermodynamics, respectively
\cite{riechersFluctuationsWhenDriving2017,semaanHomeostaticAdaptiveEnergetics2022}.
The Kullback-Leibler divergence \cite{coverElementsInformationTheory2006}
$\DKL{Z}{\Lambda}$ monitors the difference in information between the system's
state and the would-be steady state.

We refer to Eq.~\eqref{eq:IPFL_intro} as the \emph{information processing First
Law} (IPFL) since, beyond the obvious change in the joint system's information
content, the lefthand side acts as a kernel for describing a ratchet's
information processing---discussed in detail in Sec.~\ref{sec:specializations}.
Additionally, the righthand side expresses a generalized First Law used to
define excess heat and work---discussed further in
Secs.~\ref{subsec:functionals} and \ref{subsec:FL_justification}. In essence,
Eq.~\eqref{eq:IPFL_intro} expresses a First Law for the system's information
content in the same way the original equilibrium First Law does for a system's
energy (non)conservation.

Subsequently, we show that the IPFL together with stochastic thermodynamics'
integral fluctuation theorems---particularly those presented in
Refs.~\cite{hatanoSteadyStateThermodynamicsLangevin2001,semaanHomeostaticAdaptiveEnergetics2022}---generalize
and modify the two preceding asymptotic and finite-tape IPSLs. Identifying the
role of the average dissipated housekeeping heat $\expval{\Qhk}$ and the
divergence from final steady-state conditions $\DKL{Z_N}{\Lambda_N}$, it shows
that for finite-tape systems:
\begin{align}
  -\beta \expval{Q} \leq \Delta \Hent{Z} + \DKL{Z_N}{\Lambda_N} - \beta \expval{\Qhk}
  \textnormal{,}
\end{align}
where $-\expval{Q}$ corresponds to the averaged heat extracted from the thermal
environment.

With \emph{equilibrium steady states} (ESSs), this is the work $W$ done by the
ratchet-tape system: $-\expval{Q} = W$, recalling previous bounds. However, as
Sec.~ \ref{subsec:ratchets} discusses, invoking this equivalence is not
generally possible in the case of \emph{nonequilibrium steady states} (NESSs).
Instead, we give the bounds in terms of heat extracted from the thermal
environment.

Compared to the extracted-heat form of Eq.~\eqref{eq:intro_IPSL_finite}, which
gives $-\beta \expval{Q} \leq \Delta \Hent{Z}$, this explicit accounting for the
effects of NESSs and nonequilibrium \emph{dynamical} (nonsteady) start and end
configurations gives a strictly tighter bound for finite- and even-state
(defined below) systems. 

In short, for NESS systems an increase in randomness---as measured by $\Delta
\Hent{Z}$---must additionally compensate for persistent housekeeping costs---as
measured by $\beta \expval{\Qhk}$---to leverage the thermal environment as a
reservoir of extractable energy.

Finally, we demonstrate that for infinite-tape, finite-ratchet systems the
asymptotic bound is similarly tightened:
\begin{align}
  -\beta \dot{\expval{Q}} \leq \Delta h_\mu - \beta \dot{\expval{\Qhk}}
  \textnormal{,}
\end{align}
where $\beta \dot{\expval{\Qhk}}$ is the asymptotic rate of housekeeping
dissipation.

To summarize, fluctuation theorems take the IPFL directly to a suite of
simultaneously-true Second Laws for information processing. This, once again,
mirrors informational generalization of the familiar equilibrium Second Law.

Overall, this clarifies and unifies derivations of these IPSLs. More
importantly, it extends their domains to explicitly include the effects of
initial- and final-state dependence as well as nonequilibrium steady states.
Practically, this opens the door to considering detailed information-energy
tradeoffs for systems that arbitrarily violate detailed balance in their
effective dynamics. The following demonstrates this via an example ratchet
designed to tunably violate detailed balance while remaining tractable for
analysis. This uncovers qualitative corrections to a ratchet's ability to
extract heat from its environment.

The development proceeds as follows. First, Sec.~\ref{sec:prelim} sets out the
preliminary notation and introduces the relevant stochastic dynamical
functionals. Section~\ref{subsec:FL_justification} maps the general stochastic
dynamical picture to an explicitly thermodynamic one, immersed in a
single-temperature environment. This points toward concrete example realizations
of the general stochastic theory. Section~\ref{subsec:ratchets} reviews
autonomous Maxwellian information ratchets, which comprise our example system
class.

With the preliminaries in hand, Sec.~\ref{sec:ipfl_single_symbol} derives
Eq.~\eqref{eq:IPFL_intro}'s IPFL. It applies the IPFL to the information
ratchet picture, revealing strict equalities relating a ratchet's thermodynamic
dissipation with its information processing in transforming an input tape to
an output.

Section~\ref{sec:specializations} then specializes the IPFL in two ways. First,
Sec.~\ref{subsec:FTs} introduces and uses integral fluctuation theorems to take
the equality to an inequality. This arrives at the kernel of previous IPSLs,
explicitly generalizing and tightening that of
Ref.~\cite{heInformationProcessingSecond2022}. Then,
Sec.~\ref{subsec:asymptotics} considers the asymptotic rate limit of an infinite
tape, similarly generalizing the previous asymptotic IPSL to include the effects
of nonequilibrium dynamical state-dependence and potentially infinite-state
ratchets. The restriction to finite ratchets in
Sec.~\ref{subsec:finite_asymptotics} rounds out our derivations, revealing a
simple correction tightening
Ref.~\cite{boydIdentifyingFunctionalThermodynamics2016}'s asymptotic IPSL.

Finally, Sec.~\ref{sec:example_as4c} applies the developed theory to a
finite-state information ratchet that arbitrarily violates detailed balance and
so exhibits NESSs. We find that even for simple cases, NESSs have dramatic
effects on a ratchet's ability to extract heat, qualitatively changing its
landscape of allowed behaviors.

Taken together, the results (i) unify both previously-reported IPSLs for
ratchets by deriving them explicitly from the underlying IPFL and integral
fluctuation theorems, (ii) place the specific application of autonomous ratchet
function in the broader context of the exchange between energy and information
in complex systems, including generally nondetailed-balanced ratchets, and (iii)
demonstrate severe restrictions nonequilibrium dynamical states place on
allowed ratchet functionality---restrictions critical to understanding the
thermodynamics of information processing by complex systems.

\section{Preliminaries}
\label{sec:prelim}

Consider a system under study (SUS) that stochastically realizes states $z$
in a countable space $\mathcal{Z}$. It is driven in discrete time by a
\textit{protocol} written as a sequence of parameter values $\param$ in a
parameter space $\mathcal{A}$, denoted by $\param_{0:N} \defn \param_0 \param_1
\dots \param_N$ for a positive integer $N$. The resulting driven stochastic
process $Z_{0:N}$ is not stationary. However, we assume it is
\textit{conditionally stationary}: for any protocol indefinitely fixed at
$\param$, there is a unique corresponding stationary state distribution
$\bm{\pi}_\param$.

Initially, we place no further restrictions on our system: it need not have a
particular dynamical structure---Markov, master equation, Langevin, detailed
balanced, and so on. And, we make no claim about the scale of its state space
or time scale. The protocol itself may be a realization of a separate
stochastic process, and the state space may be a joint one with meaningfully
decomposable parts. In point of fact, we treat the latter as an example later.
First, though, we derive our main result in greater generality, requiring only
the conditional stationarity assumption and involving only the functionals of
trajectory-protocol pairs we now define.

\subsection{Dynamical Functionals}
\label{subsec:functionals}

With a would-be stationary distribution $\bm{\pi}_\param$ associated to each
driving parameter $\param$, denote its elements by $\pi_\param \pqty{z}$.
Without loss of generality we define the \emph{steady-state surprisal}
\cite{hatanoSteadyStateThermodynamicsLangevin2001,
riechersFluctuationsWhenDriving2017,mandalAnalysisSlowTransitions2016,
riechersFluctuationsWhenDriving2017,semaanHomeostaticAdaptiveEnergetics2022}:
\begin{align}
  \phi_\param\pqty{z} \defn - \ln \pi_\param \pqty{z}
  \textnormal{,}
\label{eq:ss_suprisal}
\end{align}
so called as it is the Shannon self-information
\cite{coverElementsInformationTheory2006} of the system state being $z$ under
the distribution $\bm{\pi}_\param$. Hereafter, we take all logarithms to
the natural base and, following information-theoretic convention, refer to the
unit of entropy and surprisal as a \emph{nat} \footnote{If the logarithm were
taken to base $2$, a \emph{bit}, and so on.}.

For notational uniformity, we cast the sequence of stationary distributions
during a protocol as a stochastic process over random variables $\Lambda_t \sim
\bm{\pi}_{\param_t}$. (``$\sim$'' reads ``distributed as'' \cite{coverElementsInformationTheory2006}.)
Upon averaging:
\begin{align}
  \braket{\bm{\pi}_\param}{\bm{\phi}_\param} & \defn 
  \sum_{z \in \mathcal{Z}} \pi_\param\pqty{z} \, \phi_\param\pqty{z} 
  \notag \\
  & = H\bqty{\Lambda}
  \textnormal{,}
\label{eq:ssd_shann_entropy}
\end{align}
the Shannon entropy of the distribution $\bm{\pi}_\param$.

We now define \emph{stochastic excess work} $\gWex$ and \emph{stochastic excess
heat} $\gQex$ as distinct contributions to a system's change in steady-state
surprisal:
\begin{alignat}{2}
  \Delta \phi &= \underbrace{\Delta_\param \phi} &&+ \underbrace{\Delta_z \phi}
    \notag \\
  &\defn \; \gWex &&- \; \gQex
  \textnormal{,}
\label{eq:1L_phi}
\end{alignat}
where, for $N$ time steps:
\begin{alignat}{2}
  \Delta_\param \phi &\defn \sum_{n=0}^{N-1} \phi_{\param_{n+1}}\pqty{z_n} -
    \phi_{\param_n} \pqty{z_n} &&\textnormal{and} \label{eq:discrete_Dparam}\\
  \Delta_z \phi &\defn \sum_{n=0}^{N-1} \phi_{\param_{n+1}}\pqty{z_{n+1}} -
    \phi_{\param_{n+1}}\pqty{z_{n}}\textnormal{.}\quad&&
  \label{eq:discrete_Dz}
\end{alignat}

That is, by stochastic excess work $\gWex$ we refer to the change in
steady-state surprisal owing to a changing environmental drive. And, by
stochastic excess heat $\gQex$ we identify the change in steady-state surprisal
owing to the system's state change---its response or \emph{adaptation} to
environmental conditions.

Next, we denote the \emph{conditional path irreversibility} by $\mathcal{Q}$:
\begin{align}
  \mathcal{Q} \defn \ln \frac{
    \cpr{Z_{1:N} = z_{1:N\phantom{-1}}}{Z_0 = z_{0\phantom{N}}; \param_{0:N}}
  }{
    \cpr{Z_{1:N} = \tilde{z}_{N-1:0}}{Z_0 = \tilde{z}_{N\phantom{0}}; \tilde{\param}_{N:0}}
  }
\textnormal{,}
\label{eq:Q_general}
\end{align}
where the tilde indicates negation of odd-parity variables, such as momentum and
magnetic field.

The stochastic excess heat $\gQex$ can be viewed as a piece of this path
irreversibility---that associated with changes in the steady-state surprisal.
What remains we term the \emph{stochastic housekeeping heat} $\gQhk$:
\begin{align}
  \gQhk \defn \mathcal{Q} - \gQex
  \textnormal{.}
\label{eq:Qhk_general}
\end{align}
In restricted cases it carries additional interpretation: if the stochastic
dynamics are Markov (order 1) \emph{and} the state and protocol variables are
even (requiring no negation in the denominator), then $\gQhk$ measures
\emph{detailed balance violation} in the stochastic dynamics. If the state or
control variables are odd, but we retain the Markov condition, then a
\emph{part} of $\gQhk$ measures detailed balance violation---see
Refs.~\cite{spinneyEntropyProductionFull2012,spinneyNonequilibriumThermodynamicsStochastic2012,yeoHousekeepingEntropyContinuous2016}
for more on this breakdown.

As stated, however, $\gQhk$ requires neither Markov dynamics nor even state and
protocol variables, and we must take care not to over-interpret. In this general
setting, it is simply that portion of a particular trajectory's
irreversibility---conditioned on initial and final configurations---that is not
attributable to changes in the system's steady-state surprisal along its forward
path.

\subsection{Nonequilibrium Dynamical States}
\label{subsec:NEDSs}

In the special case where the system is Markovian (order $1$) and subject to an
indefinitely fixed drive---yielding a stationary Markov process---the rate of
housekeeping heat takes the same form as that of asymptotic entropy production,
familiar in stochastic thermodynamics
\cite{cocconiEntropyProductionExactly2020}. The latter on average is sometimes
taken to measure the system's fundamental time-reversal asymmetry
\cite{skinnerImprovedBoundsEntropy2021,skinnerEstimatingEntropyProduction2021}.

However, the following considers the more general case of systems that have not
yet reached their steady-state distributions---processes that are not
stationary. While Eqs.~\eqref{eq:1L_phi}--\eqref{eq:Qhk_general} leverage a
suite of ``would-be'' stationary distributions, they are defined for arbitrary
paths, including when the system is nowhere near such a steady state at any
stage of the observed interval. We call such transient state distributions
$\bm{\mu}_t \nsim \bm{\pi}_{\param_t}$ \emph{nonequilibrium dynamical states}
(NEDSs). In treating system trajectories that begin and/or end in NEDSs, a final
term appearing in our derivations and related results remains: the
\emph{nonsteady-state addition to stochastic free energy}:
\begin{align}
  \Fnss_{\bm{\mu} \| \param} \pqty{z}
     \defn \ln \frac{\mu\pqty{z}}{\pi_\param \pqty{z}}
    \textnormal{.}
  \label{eq:Fnss}
\end{align}

Like $\phi$, $\gWex$, $\gQex$, $\mathcal{Q}$, and $\gQhk$, this nonsteady-state
addition to stochastic free energy's definition recalls connections to
stochastic-thermodynamic energies and entropies of interest. We have been
careful thus far to avoid over-interpretation in this vein, instead electing to
treat these quantities as stochastic-dynamically meaningful in their own right.
In the next section, however, we select a particular thermodynamic scheme and
explicitly map the preceding functionals to their thermodynamically-meaningful
counterparts.

\subsection{Thermodynamic Scheme}
\label{subsec:thermo_scheme}

\begin{figure}[ht]
\centering
\includegraphics[width=\columnwidth]{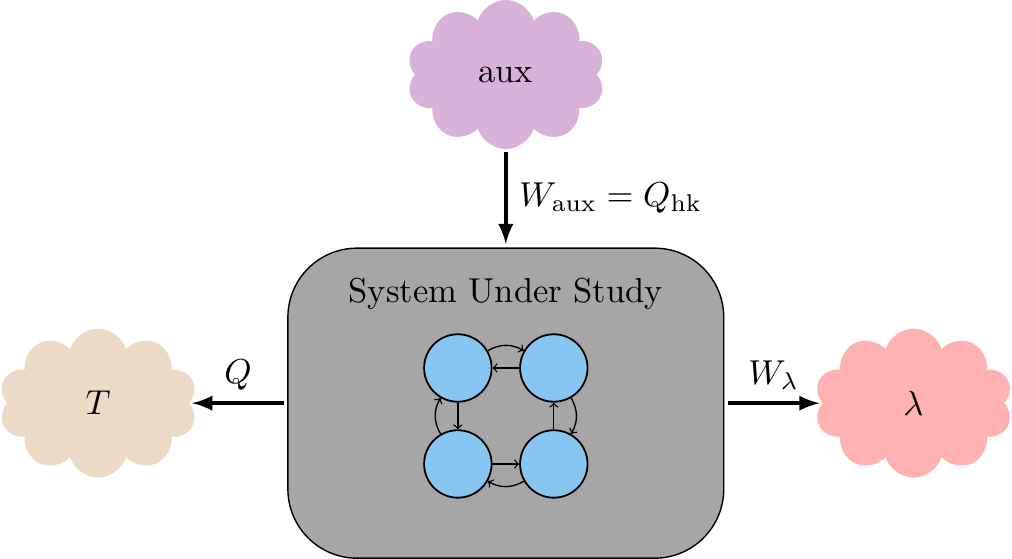}
\caption{The thermodynamic scheme considered: the Markovian stochastic system
    under study is coupled to an ideal heat bath, an ideal work reservoir
    parameterized by $\param$, and an auxiliary reservoir that accounts for
    maintaining nonequilibrium steady states. Here, we restrict the auxiliary
    reservoir's function solely to maintaining nonequilibrium steady states
    induced by the system's dynamics, so that $\abs{W_\textnormal{aux}} =
    \abs{\Qhk}$.}
\label{fig:thermo_system}
\end{figure}

To place physical constraints on our stochastically-evolving system of interest
and to aid interpretation, we now embed it in an isothermal environment at
inverse temperature $\beta \defn \kb T$, with $\kb$ Boltzmann's constant, and
connect it to two ideal \footnote{That is, exhibiting no changes in entropy.}
work reservoirs---one parameterized by $\lambda$ that couples with the
stochastic dynamics and one labeled an \emph{auxiliary reservoir}, representing
otherwise unaccounted-for degrees of freedom and providing for nonequilibrium
steady states \cite{semaanHomeostaticAdaptiveEnergetics2022}. See
Fig.~\ref{fig:thermo_system} for an illustration. Additionally, we require that
the underlying system dynamics are Markovian---or at least that there exists a
Markov chain representation of the process for each $\param$, with which we can
calculate the preceding functionals \footnote{This is certainly possible for any
finite-state Markov chain. It is not possible in general otherwise: countably
infinite-state Markov chains may not exhibit unique stationary distributions
\cite{kemenyDenumerableMarkovChains1976}, and the uncountable-state case carries
additional complications that make general extension of the functionals we
provide non-trivial.}.

If the system's dynamics are microscopic---or if the underlying coarse-graining
scheme does not include hidden entropy-producing transitions---then the
conditional path irreversibility maps directly to the total heat dissipated by
the system to its thermal environment: $\mathcal{Q} \to \beta Q$
\cite{crooksEntropyProductionFluctuation1999}. Otherwise, it provides a lower
bound for total heat---there may be unaccounted-for dissipation in hidden
degrees of freedom \cite{espositoStochasticThermodynamicsCoarse2012}. (Note
that, following our unit convention, $\beta Q$---as well as the other energies
per $\kb T$ to appear shortly---is thus measured in nats.)

Similarly, the stochastic excess and housekeeping heats map to the excess and
housekeeping heats of steady-state thermodynamics: $\gQex \to \beta \Qex$,
$\gQhk \to \beta \Qhk$, and we have $Q = \Qex + \Qhk$
\cite{riechersFluctuationsWhenDriving2017,semaanHomeostaticAdaptiveEnergetics2022,hatanoSteadyStateThermodynamicsLangevin2001,mandalAnalysisSlowTransitions2016,spinneyEntropyProductionFull2012,yeoHousekeepingEntropyContinuous2016,oonoSteadyStateThermodynamics1998,harrisFluctuationTheoremsStochastic2007}.
In this way, the total heat dissipation splits into one component due to the
system's response to environmental stimuli---the excess heat $\Qex$---and one
due to maintaining NESSs---the housekeeping heat $\Qhk$.

The nonsteady-state addition to stochastic free energy, meanwhile, becomes the
nonsteady-state (or nonequilibrium) addition to free energy, capturing
state-dependent contributions to free energy resulting from initially- (and
finally-) nonsteady-state configurations. 

Absent the ESS limit, the stochastic excess work becomes just excess work via
$\gWex \to \beta \Wex$ and carries the interpretation as that work done atop the
change in steady-state free energy that would be dissipated if the system
relaxed to its stationary distribution, given that it started in one. Or, more
specifically:
\begin{align}
  W_\textnormal{diss} \defn \Wex - \kb T
  \Delta \Fnss_{\bm{\mu} \| \param}
  \label{eq:Wdiss}
\end{align}
is the \emph{dissipated work}
\cite{riechersFluctuationsWhenDriving2017,semaanHomeostaticAdaptiveEnergetics2022}---that
done atop the change in nonequilibrium free energy as the system evolves between
two NEDSs.

Finally, underlying each of these quantities, the steady-state surprisal
$\phi_\lambda$ itself is interpreted as a \emph{nonequilibrium potential}
\cite{seifertStochasticThermodynamicsFluctuation2012}.

Unfortunately, in the NESS setting, defining a \emph{steady-state free energy}
analogous to the equilibrium free energy of ESS systems remains problematic
\cite{riechersFluctuationsWhenDriving2017,semaanHomeostaticAdaptiveEnergetics2022}.
Furthermore, directly mapping surprisals to energies is in general impossible,
as is operationally defining work and state energies
\cite{seifertStochasticThermodynamicsFluctuation2012}. As a result, we can give
explicit construction of neither $W_\lambda$ nor $W_\textnormal{aux}$.

We can, however, make useful headway by placing an additional restriction on
the auxiliary bath. Hereafter, we assume the bath provides \emph{only} that
energy required to maintain NESSs, so that $\abs{W_\param} = \abs{\Qhk}$ at all
times. This corresponds to assuming that the only unaccounted-for degrees of
freedom are those strictly necessary for maintaining nonequilibrium steady
states as implied by the observed stochastic dynamic. This means, in turn, the
ESS limit of $\gQhk \to 0$ corresponds to eliminating the auxiliary bath,
leaving a system coupled only to one ideal heat reservoir and one ideal work
reservoir, mirroring common schemes in stochastic thermodynamics
\cite{jarzynskiNonequilibriumEqualityFree1997,jarzynskiEquilibriumFreeenergyDifferences1997,boydIdentifyingFunctionalThermodynamics2016}.

\subsection{Excess and Thermodynamic First Laws}
\label{subsec:FL_justification}

We call Eq.~\eqref{eq:1L_phi} the \emph{excess First Law} due to its structural
similarity to the First Law of thermodynamics. To see why, let us now take a
detour to the $\gQhk \to 0$, ESS limit: We are left with the canonical ensemble
of statistical mechanics. Our system is affixed only to an ideal heat bath and
an ideal work reservoir and, without issue now, we assign to each state $z$ an
energy $E_\param \pqty{z}$ to obtain
\cite{jarzynskiNonequilibriumEqualityFree1997}:
\begin{alignat}{2}
  \Delta E &= \underbrace{\Delta_\param E} &&+ \underbrace{\Delta_z E} \notag \\
  &\defn \; \, W && - \; \, Q
\textnormal{.}
\label{eq:1L_energy} 
\end{alignat}
Equation~\eqref{eq:1L_energy} defines work and heat in this restricted case as
distinct contributions to the system's change in energy. Superficially,
Eq.~\eqref{eq:1L_phi} is then a First Law for steady-state surprisal in exactly
the same way that Eq.~\eqref{eq:1L_energy} is a First Law for energy.

The change of viewpoint from $E$ to $\phi$ as the central object represents a
subtle but useful generalization. Helpfully, it comes without additional risk,
since the same restrictions that give Eq.~\eqref{eq:1L_energy} also imply:
\begin{itemize}[leftmargin=*]
\item Boltzmann-distributed steady states
\cite{sethnaEntropyOrderParameters2021}, so that:
\begin{align}
  \kb T \phi_\param \pqty{z} = E_\param \pqty{z} - \Feq
  \textnormal{,}
\label{eq:noneq_potential}
\end{align}
with $\Feq$ the equilibrium free energy (the usual logarithm of the canonical
partition function);
\item For excess work \footnote{Or, equivalently, for dissipated work: 
  $
    W_\textnormal{diss} \to 
    W - \Delta F^{\textnormal{neq}} \defn 
    W - \Delta \Feq - \kb T \Delta \Fnss_{\bm{\mu} \| \param}
  $.
}:
\begin{align}
  \gWex \to \beta \pqty{W - \Delta F^{\textnormal{eq}}} 
  ~,
\label{eq:Wex_ESS}
\end{align}
we now have a consistent notion of
steady-state (equilibrium) free energy; and
\item All dissipated heat is excess:
\begin{align}
  \gQex \to \beta Q
  \textnormal{.}
\label{eq:Qex_ESS}
\end{align}
\end{itemize}

And so, for ESSs:
\begin{align}
  \kb T \Delta \phi &= \Delta E - \Delta F^{\textnormal{eq}} \notag \\
  &= \pqty{W - \Delta F^{eq}} - Q
  \label{eq:ESS_1L_mapping}
  \textnormal{.}
\end{align}
Thus, mapping from energetic to surprisal-based First Laws involves only the
switch in viewpoint from total work to \emph{excess} work as the more direct
quantity. Here, it is that work done atop the change in equilibrium free energy.
Since fluctuation theorems are phrased quite naturally in terms of functionals
of $\phi$ and realized path probabilities
\cite{hatanoSteadyStateThermodynamicsLangevin2001,
riechersFluctuationsWhenDriving2017, semaanHomeostaticAdaptiveEnergetics2022},
taking the First Law of Eq.~\eqref{eq:1L_phi} as a starting point is
particularly helpful when working with those theorems.

Treating $\phi$ as more fundamental than $E$ carries utility beyond this
convenience, however. There are many more-general settings than the canonical
ensemble. These include, for example, biological, active matter, and other NESS
systems not Boltzmann-distributed in the energies at stationarity
\cite{trepagnierExperimentalTestHatano2004,
bechingerActiveParticlesComplex2016,brownTheoryNonequilibriumFree2020}. In these
cases, in defining $\gWex$ stochastic-dynamically, Eq.~\eqref{eq:1L_phi}
circumvents issues with appropriately defining nonequilibrium steady-state free
energies
\cite{riechersFluctuationsWhenDriving2017,semaanHomeostaticAdaptiveEnergetics2022}.
Finally, in any situation where a relationship between $E$ and $\phi$ \emph{can}
be derived, one can map Eq.~\eqref{eq:1L_phi} to Eq.~\eqref{eq:1L_energy}
directly.  Moreover, the former retains its meaning and, as we shall show,
utility---even when the latter is far from familiar.

Such is often the case in highly coarse-grained, effective state-space models of
mesoscopic complex phenomena where, at best, one estimates bounds on ``true''
entropy production
\cite{rahavFluctuationRelationsCoarsegraining2007,puglisiEntropyProductionCoarse2010,espositoStochasticThermodynamicsCoarse2012,tezaExactCoarseGraining2020,ghosalInferringEntropyProduction2022}.
The coarse-grained dynamics themselves, however, may be directly observed. And
in these cases, Eq.~\eqref{eq:1L_phi} holds exactly and remains interpretable
\emph{at the level of the observed phenomena}. This is reminiscent of several
similarly-phrased fluctuation theorems; e.g.,
Ref.~\cite{semaanHomeostaticAdaptiveEnergetics2022}'s NESS trajectory class
fluctuation theorem.

\subsection{Information Ratchets}
\label{subsec:ratchets}

\begin{figure}[ht]
\centering
\includegraphics[width=\columnwidth]{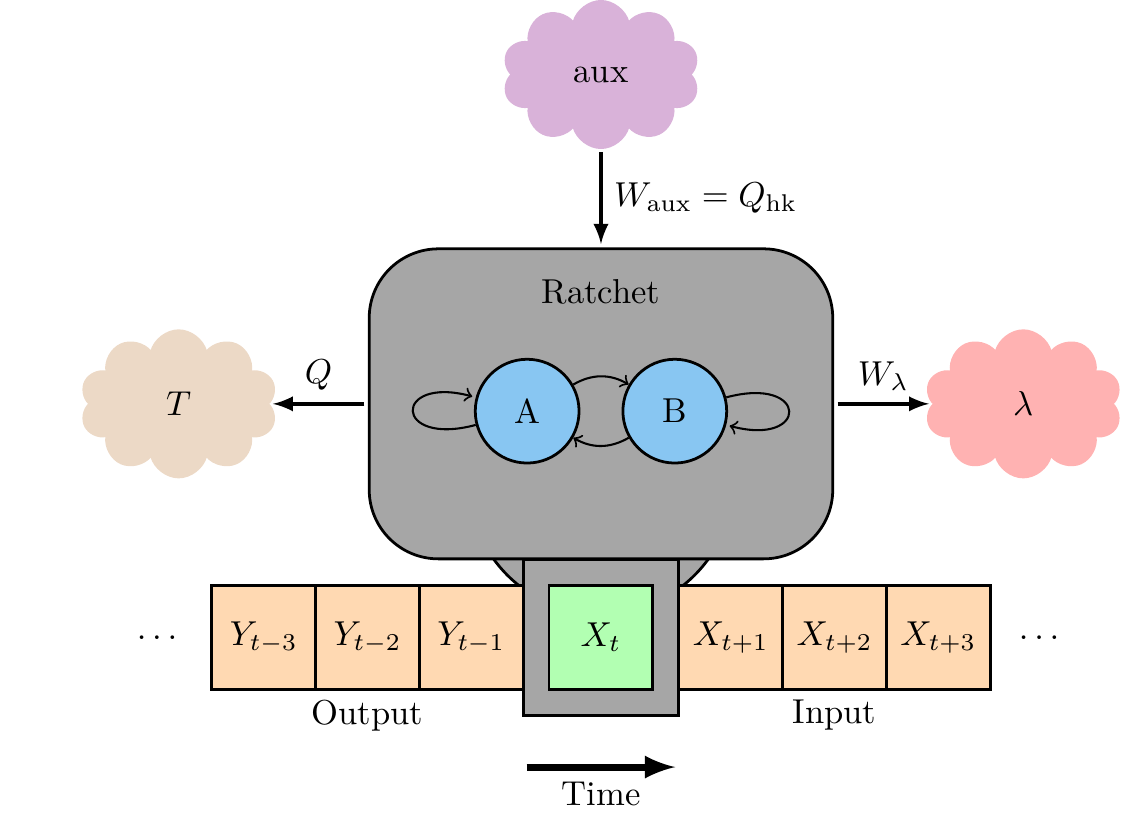}
\caption{Information ratchet system: At each time step, the ratchet moves along
  	the tape, interacting with one symbol at a time and exchanging energy with
  	the coupled reservoirs in the process. New here is the auxiliary reservoir
  	that allows for nonequilibrium steady states and another mode of energy
  	exchange with the ratchet-tape subsystem. (Illustration created in part by
  	modifying Ref.~\cite{loomisThermalEfficiencyQuantum2020}'s Fig.~1, with
  	permission from the authors.)
    }
\label{fig:ratchet_tape_system}
\end{figure}

We are especially interested in a particular decomposition of $\mathcal{Z}$ into
distinct subspaces---a \textit{ratchet} and a semi-infinite \emph{information
tape}. Figure~\ref{fig:ratchet_tape_system} illustrates the setting.

The ratchet interacts directly with only a single information-tape cell at a
time. (Hereafter, we refer to a \emph{bit} since we consider a tape with a
binary alphabet. Generalizing to other alphabet sizes is straightforward.)
Furthermore, we assume that any violation of detailed balance is strictly due to
the ratchet system interacting with a single bit. That is, there are no
energetic fluxes through the extended tape except as facilitated by the
ratchet-tape interaction. We assume that the joint dynamics of the ratchet,
interacting bit, and reservoirs is Markovian. At each time step, the ratchet:
\begin{enumerate}
\item Moves one cell along the tape, putting it in contact with the next
	interaction bit; and
\item Thermalizes (perhaps incompletely) with the coupled reservoirs for a time
	$\tau$.
\end{enumerate}

We do not assign an energy to each state in the joint dynamics, since detailed
balance is not required. (Previous studies imposed detailed balance during the
thermalization step to fix relative state energies
\cite{mandalWorkInformationProcessing2012,boydIdentifyingFunctionalThermodynamics2016,boydLeveragingEnvironmentalCorrelations2017,jurgensFunctionalThermodynamicsMaxwellian2020}.)
Indeed, in the presence of nonequilibrium steady states, assignment of state
energies---or even a formal, consistent definition of total work---is not
generally possible \cite{seifertStochasticThermodynamicsFluctuation2012}. This
precludes direct comparison with previous ratchet studies
\cite{boydIdentifyingFunctionalThermodynamics2016,jurgensFunctionalThermodynamicsMaxwellian2020,heInformationProcessingSecond2022},
wherein total work extracted was upper bounded via a lower bound on total
dissipated heat.

Rather, here we leverage the fact that in this isothermal setting $(\mathcal{Q},
\gQex, \gQhk) \to (\beta Q, \beta \Qex, \beta \Qhk)$ and consider dissipated
heats directly. In particular, when discussing thermodynamic functionality, we
focus on exchanges with the heat bath. That is, instead of the ``engine'' regime
previously defined by $\expval{W} > 0$
\cite{boydIdentifyingFunctionalThermodynamics2016,jurgensFunctionalThermodynamicsMaxwellian2020},
for example, we refer to the ``heat engine'' regime defined by $-\expval{Q} >
0$---where energy is on average extracted from the thermal reservoir.

To describe and decompose the information-bearing degrees of freedom, we split
the random variable $Z_n$ into three parts: the random variable $R_n$ (with
alphabet $\mathcal{R}_n$) corresponds to the ratchet subsystem's state at time
$n$; the joint random variable $X_{n:\infty}$ to the \emph{input tape} at time
$n$---that portion of the information tape to which the ratchet has not yet
written---and the joint random variable $Y_{0:n-1}$, for the \emph{output tape},
to which the ratchet has written. Thus, at each $n$, $Z_n = (R_n, X_{n:\infty},
Y_{0:n-1})$. This mirrors the decompositions of
Refs.~\cite{boydIdentifyingFunctionalThermodynamics2016,
jurgensFunctionalThermodynamicsMaxwellian2020}.

\section{Information Processing First Law}
\label{sec:ipfl_single_symbol}

Let us return to the general stochastic-dynamical setting, with no assumption of
any particular thermodynamic scheme. Let $\bm{\mu}_n$ be the NEDS at time step
$n$, such that $Z_n \sim \bm{\mu}_n$. Suppose we have a system that begins in
$\bm{\mu}_0$ and ends in $\bm{\mu}_N$, as driven by the protocol $\param_{0:N}$.
We begin by equating the trajectory (over all possible state-space trajectories
$z_{0:N}$) and state averages (justified in App.~\ref{apdx:avg_proof}) of the
change in steady-state surprisal:
\begin{align}
  \expval{\Delta \phi_\param} &= \Delta \braket{\mu}{\phi_\param}
  \notag \\
  &\defn \braket{\mu_N}{\phi_{\param_N}}
  - \braket{\mu_0}{\phi_{\param_0}}
    \textnormal{.}
\end{align}
The lefthand side is, by definition, $\expval{\gWex} - \expval{\gQex}$. This is
the averaged First Law for $\phi$ as in Eq.~\eqref{eq:1L_phi}.

For the righthand side, notice that:
\begin{align}
  \braket{\mu}{\phi_\param} &= \sum_{z \in \mathcal{Z}} 
  \mu\pqty{z} \phi_\param \pqty{z} \notag \\
    &= -\sum_{z\in \mathcal{Z}} \mu\pqty{z} \ln \pi_\param \pqty{z} \notag \\
    &= H\bqty{Z} + \DKL{Z}{\Lambda}
  \textnormal{.}
\label{eq:derivation_HplusD}
\end{align}
Hence, we have the \textit{information processing first law} (IPFL):
\begin{equation}
  \Delta \Hent{Z} + \Delta \DKL{Z}{\Lambda}
    = \expval{\gWex} - \expval{\gQex}
  \label{eq:EIPFL}\textnormal{.}
\end{equation}
The lefthand side accounts for the ``information processing'': the Shannon
entropy change of the system plus the change in its divergence from the local
stationary distribution. (Alternatively, the change in cross entropy between
the system's state distribution and the local steady-state distribution.)

Moving $\Delta \DKL{Z}{\Lambda}$---the averaged change in nonsteady-state free
energy from Eq.~\eqref{eq:Fnss}---to the righthand side recovers
Eq.~\eqref{eq:IPFL_intro}. This is, quite directly, a First Law for information
processing that expresses its changes in terms of averages of the
stochastic-dynamical functionals $\gWex$, $\gQex$, and $\Fnss_{\bm{\mu} \|
\param}$. These are the functionals that, under appropriate thermodynamic
schemes as in Fig.~\ref{fig:thermo_system}, carry entropic and energetic
meaning. This IPFL holds generally for transitions between NEDSs, implying
validity for NESS and even nonthermal systems, since the generalized excess
quantities are still well-defined by Eq.~\eqref{eq:1L_phi}.

Stated in the form of Eq.~\eqref{eq:EIPFL}, the IPFL makes no reference to the
``conjugate'' or ``reversed'' dynamics involved in the definitions of
$\mathcal{Q}$ and by extension $\gQhk$. Rather, it is concerned strictly with
averages weighted by forward trajectories. However, substituting
Eq.~\eqref{eq:Qhk_general} does involve these conjugated dynamics. This leads
to expressing the IPFL equivalently as:
\begin{align}
\Delta \Hent{Z} \! + \! \expval{\mathcal{Q}}
    = \expval{\gWex} \! + \! \expval{\gQhk} \! - \! \Delta \DKL{Z}{\Lambda}
    \textnormal{.}
\label{eq:EIPFL_almost_entprod}
\end{align}
Here, the lefthand side is stochastic thermodynamics' familiar (average of)
\emph{total entropy production} $\Delta S_\textnormal{tot}$, broken into system
($\Delta \Hent{Z}$) and environment ($\expval{\mathcal{Q}}$) pieces
\cite{seifertStochasticThermodynamicsPrinciples2018,riechersFluctuationsWhenDriving2017,wimsattTrajectoryClassFluctuation2022}.
The righthand side thus represents an alternative decomposition of the total
entropy production into excess environmental ($\expval{\gWex}$) and housekeeping
($\expval{\gQhk}$) components, as well as one due to initial (and final) state
dependence ($\Delta \DKL{Z}{\Lambda}$)
\cite{riechersFluctuationsWhenDriving2017,kolchinskyDependenceDissipationInitial2017,
riechersInitialstateDependenceThermodynamic2021,
semaanHomeostaticAdaptiveEnergetics2022}. The IPFL, then, expresses a particular
decomposition of the average total entropy production. In the appropriate
settings, the decomposition directly links change in information content with
thermodynamic processes \emph{without} invoking conjugated dynamics.

\subsection{Application to Information Ratchets}

Arriving at Eq.~\eqref{eq:EIPFL} required minimal assumptions about the
underlying SUS. Now, we wish to specialize it to the information ratchet system
of Sec.~\ref{subsec:ratchets} and Fig.~\ref{fig:ratchet_tape_system}. In
particular, the isothermal environment takes our stochastic-dynamical
functionals to thermodynamic energies. And, the distinct ratchet and tape
subspaces allow for meaningful decomposition of the information-bearing degrees
of freedom.

First, we expand $\Delta \Hent{Z} = \Hent{Z_N} - \Hent{Z_0}$ from
Eq.~\eqref{eq:EIPFL}. Splitting the joint Shannon entropies, making use of
mutual informations---denoted $\Imut{\cdot}{\cdot}$ for the (symmetric) mutual
information between two random variables
\cite{coverElementsInformationTheory2006}---and bearing in mind that changes in
indices for $X, Y, R$ are to be inferred from the breakdowns of $Z_N$ and $Z_0$:
\begin{align}
  \Delta \Hent{Z} &= \Delta \Hent{R} 
    + \Delta \Hent{X, Y} - \Delta \Imut{R}{X,Y} \notag \\
  &= \Delta \Hent{R} + \Delta \Hent{X} + \Delta \Hent{Y} \notag \\
  &\qquad - \Delta \Imut{R}{X,Y} - \Delta \Imut{X}{Y} 
  \label{eq:DH_terms}
  \textnormal{.}
\end{align}
This further decomposes the IPFL of Eq.~\eqref{eq:EIPFL}:
\begin{align}
  \expval{\gWex} &- \expval{\gQex} \notag \\ 
  & = \Delta \Hent{R} + \Delta \Hent{X} + \Hent{Y_{0:N-1}} \notag \\
  &\qquad - \Delta \Imut{R}{X,Y} - \Imut{X_{N:\infty}}{Y_{0:N-1}} \notag \\
  &\qquad + \Delta \DKL{Z}{\Lambda}
  \textnormal{.}
\label{eq:EIPFL_DH_decomposed}
\end{align}
Equivalently, the decomposition of average total entropy production in
Eq.~\eqref{eq:EIPFL_almost_entprod} becomes:
\begin{align}
  \expval{\gWex} & + \expval{\gQhk} - \Delta \DKL{Z}{\Lambda} \notag \\
  & = \expval{\mathcal{Q}} + \Delta \Hent{R}
  + \Delta \Hent{X} + \Hent{Y_{0:N-1}} \notag \\
  &\qquad - \Delta \Imut{R}{X,Y} - \Imut{X_{N:\infty}}{Y_{0:N-1}}
  \textnormal{.}  
  \label{eq:EIPFL_DH_decomposed_entprod}
\end{align}

Equation~\eqref{eq:DH_terms}'s decomposition took Eqs.~\eqref{eq:EIPFL} and
\eqref{eq:EIPFL_almost_entprod} to Eqs. \eqref{eq:EIPFL_DH_decomposed} and
\eqref{eq:EIPFL_DH_decomposed_entprod}, respectively. The decomposition is
identical to that in Ref.~\cite{boydIdentifyingFunctionalThermodynamics2016}.
However, there the goal was to take asymptotic rates. That, together with the
finite-state ratchet requirement, removed several terms. Here, we pause to
interpret each term in its finite-time context and comment on its contribution
to the averaged total entropy production.

The first term $\expval{\mathcal{Q}}$ is the environment's contribution to the
total entropy production. All that remains is contributed by the joint
ratchet-tape system.

The second term $\Delta \Hent{R}$ monitors the change in information content of
the ratchet's states---a change in the ratchet's internal memory. If, as the
ratchet interacts with the tape, it gains memory in this sense, this specific
part of the joint system must become more randomized. And, equivalently, this
term contributes an increase to the total entropy production.

The third term $\Delta \Hent{X} = \Hent{X_{N:\infty}} - \Hent{X_{0:\infty}}$
quantifies a change in the information content of the input tape. Or, more
specifically for finite alphabets, this is strictly nonpositive---the opposite
of the information contributed by the random variables $X_{0:N-1}$. And so, the
more random the input tape, the more negative this term can be. We expect
memoryless inputs to reduce the potential to extract heat compared to memoryful
ones. That is, colloquially there is less pattern to scramble
\cite{jurgensFunctionalThermodynamicsMaxwellian2020}. We shall see later that
this is indeed the case for IPSLs. For the IPFL, in the meantime, this term's
negativity reduces averaged total entropy production. Intuitively, removing
randomness in the input tape reduces overall entropy production.

The fourth term $\Delta \Hent{Y}$ is the output tape's information content.
Its impact on countable spaces---as assumed---is straightforward. Due to
Shannon information's nonnegativity, the more random the ratchet makes the
output tape, the greater the positive contribution to average total entropy
production.

The fifth term $\Delta \Imut{R}{X,Y}$ tracks the change in \emph{shared}
information between the ratchet and tape. As the ratchet interacts with bits
from the input tape and writes to the output tape, it induces correlation
between it and the tape. While at first glance this recalls
Ref.~\cite{jurgensFunctionalThermodynamicsMaxwellian2020}'s (de)randomizer
axes, it is altogether different. Those axes tracked induced correlations
\emph{internal} to the information tape, whereas this term tracks induced
correlation \emph{between the ratchet and the (entire input-output) tape}.
Mutual information's positivity makes the contribution to
Eq.~\eqref{eq:DH_terms} strictly nonpositive, lowering the average total
entropy production. In other words, inducing correlation between the ratchet
and tape reduces the joint system's entropy production and vice versa.

This mutual information is especially important to consider in tandem with
$\Delta \Hent{R}$. While an increase in ratchet memory alone---unrelated to tape
correlation---acts to increase entropy production, generically an increase in
ratchet memory also enables greater correlation with the information tape via
$\Delta \Imut{R}{X,Y}$. Thus, the ratchet memory's effect on entropy production
involves both terms: correlation between the ratchet and tape, enabled by
ratchet memory to capture temporal patterns in the tape, counteracts the entropy
produced by an increase in ratchet memory alone.

Finally, the sixth term $\Imut{X_{N:\infty}}{Y_{0:N-1}}$ is the
mutual information between the input and output tapes. And so, again due to
mutual information's nonnegativity, it has the effect of decreasing the averaged
total entropy production.

Taken altogether, Eqs.~\eqref{eq:EIPFL_DH_decomposed} and
\eqref{eq:EIPFL_DH_decomposed_entprod} delineate exact links between finite-time
ratchet-tape information processing and the joint system's thermodynamic
behavior. It is a specialization of the IPFL to the case of a system constructed
as in Fig.~\ref{fig:ratchet_tape_system}. Shortly, we use it as a starting point
to derive and generalize the previously-reported IPSLs for ratchet-tape systems
\cite{boydIdentifyingFunctionalThermodynamics2016,heInformationProcessingSecond2022}.
However, the inequalities in IPSLs are replaced by equalities of the IPFL in the
same way that fluctuation theorems of stochastic thermodynamics replace
inequalities of thermodynamic Second Laws with strict equalities. In point of
fact, as we now show, fluctuation theorems directly take the IPFL to IPSLs.

\section{Ratchet First to Second Laws}
\label{sec:specializations}

The following derives several specializations to the information ratchet system
class of Sec.~\ref{subsec:ratchets} and Fig.~\ref{fig:ratchet_tape_system},
starting from Eqs.~\eqref{eq:EIPFL} and \eqref{eq:EIPFL_almost_entprod}. To do
this, it first leverages an integral fluctuation relation to take the equality
to an inequality. It then splits the effective state space as in
Sec.~\ref{subsec:ratchets}, along the way generalizing part of a
recently-reported finite-tape IPSL \cite{heInformationProcessingSecond2022}.
Finally, it takes asymptotic limits to similarly generalize the previous
asymptotic IPSL
\cite{boydIdentifyingFunctionalThermodynamics2016,boydLeveragingEnvironmentalCorrelations2017,jurgensFunctionalThermodynamicsMaxwellian2020}
for these regimes. Since we adopt the same assumptions as
Sec.~\ref{subsec:ratchets}, hereafter the underlying dynamics of the joint
ratchet-tape space are Markovian. However, the statistical process that produces
input and output tape symbol sequences need not be Markovian. In the infinite
tape case, they may even possess infinite-range temporal correlations.

\subsection{Fluctuations and Second Laws}
\label{subsec:FTs}

A crowning achievement of stochastic thermodynamics over the last several
decades was the development fluctuation relations and
fluctuation theorems that capture fluctuations arbitrarily far from
equilibrium. (See Ref.~\cite{seifertStochasticThermodynamicsPrinciples2018} for
a recent review.) These come in three main types: (i) \emph{integral}, relating
to exponential averages over all possible trajectories
\cite{jarzynskiEquilibriumFreeenergyDifferences1997,jarzynskiNonequilibriumEqualityFree1997,hatanoSteadyStateThermodynamicsLangevin2001,speckIntegralFluctuationTheorem2005};
(ii) \emph{detailed}, exposing a time reversal (a)symmetry between forward and
reverse paths
\cite{crooksEntropyProductionFluctuation1999,crooksNonequilibriumMeasurementsFree1998,riechersFluctuationsWhenDriving2017,mandalEntropyProductionFluctuation2017,lahiriFluctuationTheoremsExcess2014};
and \emph{trajectory class}, interpolating between the two
\cite{wimsattHarnessingFluctuationsThermodynamic2021,
semaanHomeostaticAdaptiveEnergetics2022, wimsattTrajectoryClassFluctuation2022}.

A comprehensive review would go too far afield here; rather see Refs.
\cite{yangUnifiedFormalismEntropy2020,
harrisFluctuationTheoremsStochastic2007}. Nonetheless, the following uses, in
particular, two integral fluctuation theorems:
\begin{align}
  1 &= \expval{\ex^{-\pqty{ \Delta S_\textnormal{tot}}}}
    \qquad \textnormal{and} 
    \label{eq:total-IFT}\\
  1 &= \expval{\ex^{-\pqty{ \gWex + \Fnss_{\bm{\mu}_0 \| \param_0 }}}}
    \textnormal{.} 
\label{eq:HT-IFT_generalized}
\end{align}

Invoking the convexity of the exponential, we apply Jensen's inequality to
derive the \emph{generalized Second Laws}:
\begin{align}
  \Delta \Hent{Z} + \expval{\mathcal{Q}} & \geq 0
  \qquad \qquad \textnormal{ and}
  \label{eq:tot-2L_gen} \\
  \expval{\gWex} + \DKL{Z_0}{\Lambda_0} & \geq 0
  \textnormal{.}
\label{eq:Wex-2L_gen}
\end{align}
Equation~\eqref{eq:tot-2L_gen} is thus a consequence of the total entropy
production integral fluctuation theorem. Equation
\eqref{eq:HT-IFT_generalized}, first introduced in
Ref.~\cite{semaanHomeostaticAdaptiveEnergetics2022}, generalizes the integral
fluctuation theorem of Ref. \cite{hatanoSteadyStateThermodynamicsLangevin2001}
to include initial-state dependence. The resulting inequality in
Eq.~\eqref{eq:Wex-2L_gen} shows that initially-nonsteady states \emph{lower}
the bound on $\gWex$. 

The following now demonstrates a suite of IPSLs that result directly from
applying these integral fluctuation theorems and Jensen's inequality to the
IPFL.

First, Eq.~\eqref{eq:tot-2L_gen} gives, directly:
\begin{align}
  -\expval{\mathcal{Q}} \leq \Delta \Hent{Z}
  \label{eq:Q_IPSL_totIFT}
  \textnormal{.}
\end{align}
While there was no need to substitute into an IPFL expression, note that
Eq.~\eqref{eq:tot-2L_gen}'s lefthand side is identical to
Eq.~\eqref{eq:EIPFL_almost_entprod}'s righthand side, and so
Eq.~\eqref{eq:Q_IPSL_totIFT} is a specialization of that equality.

Notice that $\expval{\mathcal{Q}}$ can be negative so long as $\Delta \Hent{Z}$
is positive. In an appropriate thermal environment, such as that of
Fig.~\ref{fig:ratchet_tape_system}, this upper bounds the finite-time extracted
heat and it is (the negation of)
Ref.~\cite{boydIdentifyingFunctionalThermodynamics2016}'s Eq.~(A7). This is at
the core of the latter's subsequent derivation, as it shows how energy may be
extracted from a heat bath at the cost of an increase in the system's
information-bearing entropy.

In the ESS ratchet setting, such as that considered by
Refs.~\cite{boydIdentifyingFunctionalThermodynamics2016,jurgensFunctionalThermodynamicsMaxwellian2020,heInformationProcessingSecond2022},
we may also rephrase this bound in terms of the averaged work $\beta \expval{W}$
done \emph{by} the system \footnote{That is, $\beta \expval{\Delta E} = 0 =
-\beta \expval{W} - \expval{\mathcal{Q}}$. The signs here indicate that heat
flowing from the bath into the ratchet is done by the ratchet on its nonthermal
surroundings.}:
\begin{align}
  \beta \expval{W} \leq \Delta \Hent{Z}
  \label{eq:W_IPSL_totIFT}
  \textnormal{.}
\end{align}
This is exactly Ref.~\cite{heInformationProcessingSecond2022}'s finite-tape,
single-pass IPSL, where $Z$ denotes the joint random variable of their ratchet
and tape subspaces.

However, in the NESS setting one cannot directly equate negative dissipated
heat with positive work production, as detailed previously in
Sec.~\ref{subsec:ratchets}. To avoid confusion, then, we focus on
Eq.~\eqref{eq:Q_IPSL_totIFT}'s heat bound: Negative averaged total heat
indicates the system's function as a \emph{heat engine}, a net extraction of
energy from the thermal environment.

The NESS setting, however, affords us additional IFTs. We can, for example,
substitute Eq.~\eqref{eq:HT-IFT_generalized} into Eq.~\eqref{eq:EIPFL},
yielding:
\begin{align}
  -\expval{\gQex} \leq \Delta \Hent{Z} + \DKL{Z_N}{\Lambda_N}
  \label{eq:Qex_IPSL_kernel}
\end{align}
or, equivalently, via Eq.~\eqref{eq:Qhk_general}:
\begin{align}
  -\expval{\mathcal{Q}} \leq 
    \Delta \Hent{Z} + \DKL{Z_N}{\Lambda_N}
    - \expval{\gQhk}
    \textnormal{.}
    \label{eq:Q_IPSL_kernel}
\end{align}
This adjusts the finite-time extracted heat bound to account for NEDSs. First,
for finite spaces, nonsteady final states raise the bound on extracted heat.
That is, we need not dissipate to full relaxation. The presence of NESSs instead
\emph{lowers} the bound by the amount of the total housekeeping heat. Thus, as
long as $\gQhk \geq \DKL{Z_N}{\Lambda_N}$---the result of an established IFT in
even state spaces
\cite{semaanHomeostaticAdaptiveEnergetics2022}---Eq.~\eqref{eq:Q_IPSL_kernel} is
always a tighter bound than Eq.~\eqref{eq:Q_IPSL_totIFT}.

In this way, Eq.~\eqref{eq:Q_IPSL_kernel} reveals the effect of NEDSs on the
heat-bath equivalent of Ref.~\cite{heInformationProcessingSecond2022}'s
Eq.~(19) and establishes it under very general conditions. Additionally, two
new effects appear. The ensemble-averaged housekeeping heat $\gQhk$
\emph{lowers} the bound on heat extraction---as an additional source of
dissipation---while the final-state dependence $\DKL{Z_N}{\Lambda_N}$ raises
it. That is, we need not account for what \emph{would be} dissipation if the
system fully relaxed to its steady states
\cite{kolchinskyDependenceDissipationInitial2017}. As before, this bound is
always tighter than Eq.~\eqref{eq:W_IPSL_totIFT} in the case of even state
spaces.

Finally, applying the preceding decomposition of $\Delta \Hent{Z}$ to
Eq.~\eqref{eq:Q_IPSL_kernel} gives the analogue to
Ref.~\cite{heInformationProcessingSecond2022}'s finite-tape IPSL, but further
decomposed to both account for NEDSs and delineate ratchet-tape information
dynamics:
\begin{align}
  -\beta \expval{Q} &\leq \DKL{Z_N}{\Lambda_N} - \expval{\gQhk}
  \notag \\
  & \quad + \Delta \Hent{R} + \Delta \Hent{X} + \Hent{Y_{0:n-1}} \notag \\
  & \quad - \Delta \Imut{R}{X, Y} - \Imut{X_{N:\infty}}{Y_{0:N-1}}
  \textnormal{.} \label{eq:IPSL-Q-DHdecomposed}
\end{align}
With this, we can translate how each term of $\Delta \Hent{Z}$ affected the
averaged total entropy production to its effect on the maximum extracted work.
In short, information processing that reduces the averaged total entropy
production identically reduces the upper bound on ensemble-averaged heat
extraction. That is, even for NEDS, extracting heat requires producing entropy.

\subsection{General Asymptotics}
\label{subsec:asymptotics}

The preceding results apply for all finite times or, equivalently, for finite
tapes. Now, we address asymptotics. Our procedure is to take $N\to\infty$
and divide the quantities of interest by $N$, giving an asymptotic rate per time
step. For notational simplicity, we use the dot notation for the thermodynamic
quantities:
\begin{align}
  \dot{\expval{\mathcal{Q}}} \defn 
    \lim_{N\to\infty} \frac{1}{N} \expval{\mathcal{Q}}
\end{align}
and so on, for $\dot{\expval{\gWex}}$, $\dot{\expval{\gQex}}$,
$\dot{\expval{\gQhk}}$, and $\dot{\expval{W}}$.

Let's take the asymptotic limit of Eq.~\eqref{eq:EIPFL_DH_decomposed}. In
particular, as in Ref.~\cite{boydIdentifyingFunctionalThermodynamics2016} we
have (i) $\lim_{N\to\infty} \Delta \Hent{X} / N = -h_\mu$, (minus) the Shannon
entropy rate of the process generating the input tape; (ii) $\lim_{N\to\infty}
\Hent{Y_{0:N-1}}/N = h'_\mu$, the Shannon entropy rate of the process
generating the output tape; and (iii) $\lim_{N\to\infty}
\Imut{X_{N:\infty}}{Y_{0:N-1}}/N = 0$.

The remaining two pieces of $\Delta \Hent{Z}$, however, vanish only under
restricting to finite-state ratchets. Without that assumption we are left with
an asymptotic IPFL:
\begin{align}
  &\dot{\expval{\gWex}} - \dot{\expval{\gQex}} = \Delta h_\mu \notag \\
     &\qquad \qquad \quad + \lim_{N\to \infty} \frac{1}{N} \pqty{
    \Delta \Hent{R} - \Delta \Imut{R}{X, Y}
  } \notag \\
  &\qquad \qquad \quad + \lim_{N\to \infty} \frac{1}{N}\Delta \DKL{Z_N}{\Lambda_N}
  \label{eq:EIPFL_asymptotic}
  \textnormal{.}
\end{align}
And, similarly, write the heat-extraction asymptotic IPSL:
\begin{align}
  &-\beta \dot{\expval{Q}}  \leq \Delta h_\mu - \dot{\expval{\gQhk}}
    \notag \\
  &\qquad \qquad \quad + \lim_{N\to \infty} \frac{1}{N} \pqty{
    \Delta \Hent{R} - \Delta \Imut{R}{X, Y}
  } \notag \\
  &\qquad \qquad \quad + \lim_{N\to \infty} \frac{1}{N} \DKL{Z_N}{\Lambda_N}
  \textnormal{.}
\label{eq:IPSL_asymptotic}
\end{align}
This generalizes the previous bound by accounting explicitly for
final-state dependence, nonequilibrium steady states, and potentially
infinite-state ratchets.

We leave detailed analytical consideration of the remaining limits for
infinite-state ratchets and their con/divergence to a sequel. However, we will
interpret the contextual meaning of the remaining limits for countably infinite
ratchets.

First, $\lim_{N\to\infty} \Delta \Hent{R}/N$ is the rate of change of the
ratchet's \emph{statistical complexity} $C_\mu \bqty{R}$ per time step, lower
bounded by the statistical complexity of its $\epsilon$-machine representation
from computational mechanics
\cite{jurgensDivergentPredictiveStates2021,crutchfieldOrderChaos2012}. In
essence, this limit measures the rate of increase in ratchet memory as it reads
an infinite stream of incoming bits. It is only nonzero for a ratchet with an
infinite memory capacity. The resulting device is able to violate the
finite-state asymptotic IPSL by leveraging its infinite internal memory to
produce work in excess of that bound
\cite{boydLeveragingEnvironmentalCorrelations2017}. For any finite-state
ratchet $\lim_{N\to\infty} \Delta \Hent{R}/N$ vanishes since in that case
$\Hent{R_N}$ is bounded from above. It vanishes also for any infinite-state
ratchet that does not asymptotically \emph{gain} memory from an infinite stream
of inputs. More precisely, this holds for a ratchet whose internal state
distribution approaches a fixed value unaffected by the incoming bit stream.

Second, $\lim_{N\to\infty} \Delta \Imut{R}{X,Y}/N$ is the rate of change of
correlation between the ratchet and the total information tape. This limit
is nonzero only if (i) the ratchet continually gains memory as above and
(ii) the ratchet continually induces correlation between itself and the total
input-output tape.

Finally, $\lim_{N\to\infty} \Delta \DKL{Z}{\Lambda}/N$ monitors (asymptotic)
movement away from steady-state conditions. Specifically, it is nonzero only
when $\DKL{Z_N}{\Lambda_N}$ diverges with $N$---the system approaches a
distribution infinitely far from stationarity---perhaps by moving progressively
farther away from the underlying stationary distribution at each step.
Colloquially, the interaction timescale is so short that the ratchet-tape system
moves further away from thermalization by constantly changing the interaction
bit. The presence of this term at all implies the existence of a stationary
distribution for the ratchet-tape system. This is a fact not guaranteed for
infinite ratchets
\cite{boydLeveragingEnvironmentalCorrelations2017,kemenyDenumerableMarkovChains1976},
but assumed by our stochastic (thermo)dynamical formalism.

\subsection{Finite Ratchet Asymptotics}
\label{subsec:finite_asymptotics}

Assuming a finite-state ratchet---in line with potential physical
implementation---simplifies the asymptotic analysis. (As it did in
Refs.~\cite{boydIdentifyingFunctionalThermodynamics2016,
boydLeveragingEnvironmentalCorrelations2017}.) This results in an asymptotic
IPFL for finite-state systems:
\begin{align}
  \dot{\expval{\gWex}} - \dot{\expval{\gQex}} = 
    \Delta h_\mu
  \textnormal{.}
\label{eq:asymptotic_IPFL_simplified}
\end{align}
And, finally, there is the correction to the previously-reported
IPSL---rewritten as a bound on extracted heat---for finite ratchets interacting
with an infinite tape:
\begin{align}
  -\beta \dot{\expval{Q}} \leq \Delta h_\mu - \dot{\expval{\gQhk}}
  \label{eq:asymptotic_IPSL_simplified}
  \textnormal{.}
\end{align}
The correction is simply $\beta \dot{\expval{\Qhk}}$ in the isothermal setting.
For even state spaces this is nonnegative and so tightens the previous bound.
Said simply, housekeeping dissipation reduces the maximum extracted heat---one
cannot harness what must go toward maintaining NESSs.

\section{Asymmetric Stochastic 4-Cycle}
\label{sec:example_as4c}

The presence of housekeeping dissipation in
Eq.~\eqref{eq:asymptotic_IPSL_simplified} suggests meaningful change in ratchet
functional thermodynamics \cite{Boyd15a}, depending on the degree to which the
joint ratchet-bit system violates detailed balance. To demonstrate this
dependence we introduce the \emph{asymmetric stochastic $4$-cycle} (AS4C)---a
two-state ratchet coupled to an information tape. The states are labeled
$\mathrm{A}$ and $\mathrm{B}$. With joint ratchet-bit Markov chain states
ordered by $\pqty{\mathrm{A}\otimes 0, \mathrm{A} \otimes 1, \mathrm{B} \otimes
1, \mathrm{B} \otimes 0}$, the dynamics obey the row-stochastic transition matrix:
\begin{align}
  \mathbf{T}\pqty{p,q} \defn 
  \begin{bmatrix}
    0 & p & 0 & 1-p \\
    1-p & 0 & p & 0 \\
    0 & 1-qp & 0 & qp \\
    p & 0 & 1-p & 0
  \end{bmatrix}
  \label{eq:T_as4c}
  \textnormal{.}
\end{align}

\begin{figure}[ht]
\centering
\includegraphics{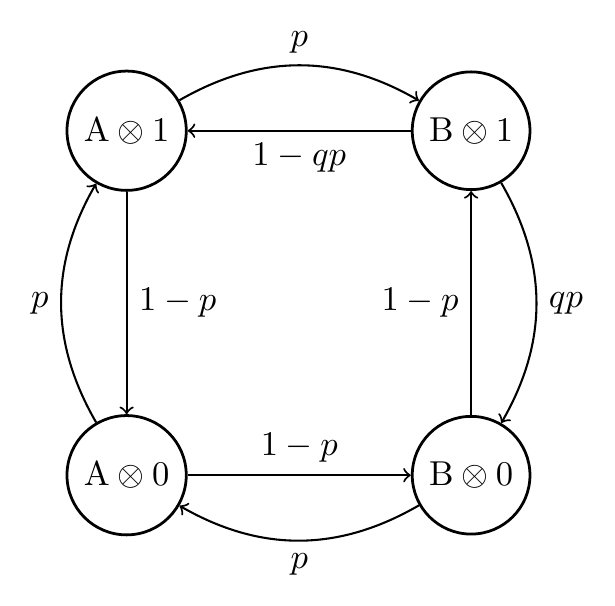}
\caption{Markov chain describing the joint ratchet-bit dynamics of the
	asymmetric stochastic 4-cycle (AS4C) ratchet family. The behavior is
	parameterized by $p \in \left( 0,1\right)$ and $q \in \left( 0,1\right]$.
	The ratchet generically violates detailed balance. When $q \neq 1$ it
	controls rotational asymmetry and can probabilistically favor either $0 \to
	1$ or $1 \to 0$ transitions. The symmetric value $q = 1$ equally favors
	these transitions, but the dynamics of the joint space still exhibits
	directionality. Detailed balance is satisfied in this case only for $p =
	1/2$.
	}
\label{fig:as4c_mc}
\end{figure}

This two-parameter ratchet family, pictured in Fig.~\ref{fig:as4c_mc},
generically violates detailed balance and allows $0 \to 1$ and $1 \to 0$
transitions to be unequally favored in terms of transition probabilities. The
latter fact manifests as a rotational asymmetry in the cycle---given by the
scaling parameter $q \in \left( 0, 1 \right]$. When $q = 1$, the cycle is
symmetric: $0 \to 1$ and $1 \to 0$ transitions are equally favored, but the
system exhibits stationary directionality in its joint state space. In the
symmetric case, detailed balance is achieved only when $p = \frac{1}{2}$.

The extent to which a discrete- and even-state Markov chain system violates
detailed balance on average is given by $\dot{\expval{\gQhk}}$. Via a
single-step average of Eq.~\eqref{eq:Qhk_general} we thus obtain for discrete
time:
\begin{align}
  \dot{\expval{\gQhk}}
  = \sum_{i \neq j} \pi \pqty{i} \bqty{\mathbf{T}}_{ij} \log
    \frac{
      \pi \pqty{i} \bqty{\mathbf{T}}_{ij}
    }{
      \pi \pqty{j} \bqty{\mathbf{T}}_{ji}
    }
    \textnormal{,}
    \label{eq:avg_dqhk}
\end{align}
where $i$ and $j$ index the states.

\begin{figure}[ht]
\centering
\includegraphics{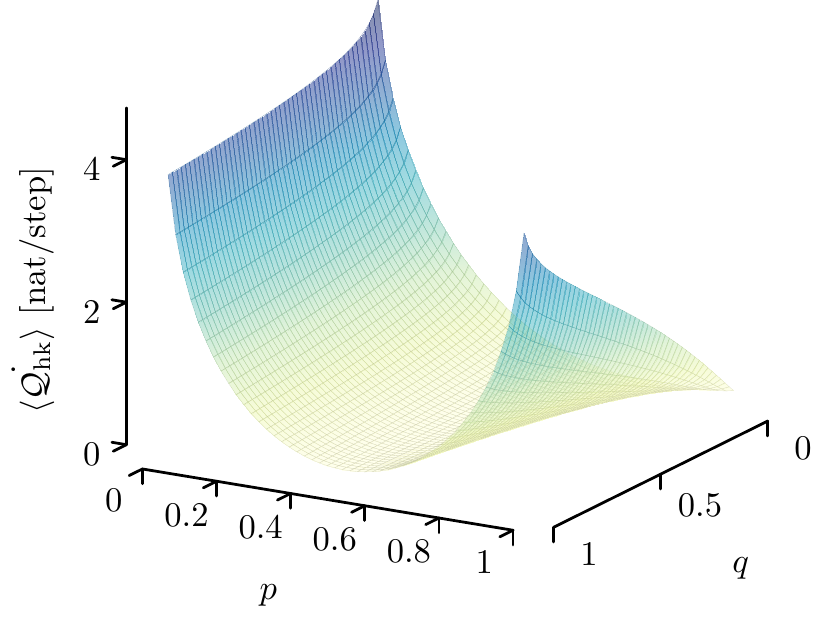}
\caption{Averaged rate of housekeeping entropy production
	$\dot{\expval{\gQhk}}$---measured in nats per time step, in keeping with our
	convention that surprisal takes units of nat---for the asymmetric stochastic
	$4$-cycle, as a function of parameters $p$ and $q$. This is the exact amount
	by which Eq.~\eqref{eq:asymptotic_IPSL_simplified} tightens
	Eq.~\eqref{eq:alec_IPSL}.
  }
\label{fig:AS4C_dqhk}
\end{figure}

This is also the exact amount by which the previous asymptotic ESS IPSL
Eq.~\eqref{eq:alec_IPSL} was tightened by our NESS IPSL in
Eq.~\eqref{eq:asymptotic_IPSL_simplified}. To visualize this difference---the
degree of tightening---Fig.~\ref{fig:AS4C_dqhk} plots $\dot{\expval{\gQhk}}$
while sweeping parameters $p$ and $q$.

\subsection{Input-Output Transducer}
\label{subsec:transducer}

As one sees, $\dot{\expval{\gQhk}}$ is far from zero over a wide range of the
parameter space. These are entropies that must be produced---equivalently in the
isothermal setting, heat that must be dissipated---to maintain the system's NESS
character. And so, one expects, they significantly impact the system's ability
to leverage an information reservoir to extract heat from a thermal environment.
This is to say, with our correction to the ESS IPSL in hand, we can analyze
bounds on the functional thermodynamics of this ratchet family.

To do so, we must calculate the remaining terms in
Eq.~\eqref{eq:asymptotic_IPSL_simplified}, namely the Shannon entropy rates
$h_\mu$ and $h'_\mu$ of the processes generating the input and output state
sequences. Following
Refs.~\cite{boydIdentifyingFunctionalThermodynamics2016,jurgensFunctionalThermodynamicsMaxwellian2020},
we achieve this by first translating our $4$-state joint ratchet-bit Markov
chain into a $2$-state ratchet \emph{transducer} that accepts as input the
process generating the input symbol statistics---in the form of a hidden Markov
chain (HMC)---and produces as output the HMC generating the output symbol
statistics.

A transducer is specified by its input-output-labeled matrices
$\Mgen$:
\begin{align}
  \Mgen \defn \mathbb{P}^{\mathsf{T}}_\textnormal{in}
    \, \mathbf{T} \, 
    \mathbb{P}^{\phantom{\mathsf{T}}}_\textnormal{out}
    \label{eq:Mio_defn}
    \textnormal{.}
\end{align}
The AS4C ratchet has two projection matrices $\mathbb{P}_0$ and $\mathbb{P}_1$
given by:
\begin{align}
  \mathbb{P}_0 = 
  \begin{bmatrix}
    1 & 0 \\
    0 & 0 \\
    0 & 0 \\
    0 & 1
  \end{bmatrix}
  \quad \textnormal{and} \quad 
  \mathbb{P}_1 = 
  \begin{bmatrix}
    0 & 0 \\
    1 & 0 \\
    0 & 1 \\
    0 & 0
  \end{bmatrix}
  \label{eq:AS4C_projectors}
  \textnormal{.}
\end{align}

\begin{figure}[ht]
\centering
\includegraphics{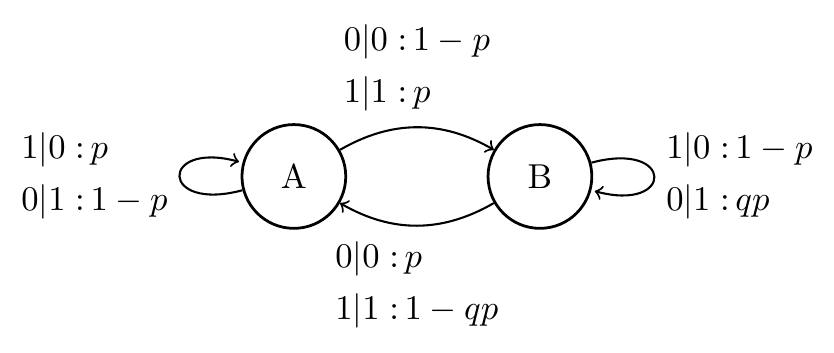}
\caption{The transducer corresponding to the $4$-state joint ratchet-bit
  Markov chain of the asymmetric stochastic $4$-cycle (AS4C). The transducer has
  two internal states, A and B, and accepts $0$ or $1$ as an input bit symbol.
  The transitions (graph edges) are labeled $\textnormal{output bit} |
  \textnormal{input bit}: \textnormal{probability}$.
  }
\label{fig:as4c_trans}
\end{figure}

This defines the AS4C's transducer, whose state-transition diagram is
visualized in Fig.~\ref{fig:as4c_trans}. Now, we compose it with any input
HMC---specified by its symbol-labeled transition matrices
$\mathbf{U}^\pqty{x}$---to give the output HMC producing the symbol statistics
on the output tape, specified by $\mathbf{V}^\pqty{y}$
\cite{barnettComputationalMechanicsInput2015,
jurgensFunctionalThermodynamicsMaxwellian2020}. The output HMC state space is
the Cartesian product of the state spaces of the input HMC and the transducer.
Let $i$ and $j$ index the states of the input HMC and $i'$ and $j'$ index those
of the transducer. Then: 
\begin{align}
  V^\pqty{y}_{i \cross i' \to j \cross j'} = 
  \sum_x M^{\left( y \middle| x \right)}_{i'j'} U^\pqty{x}_{ij}
  \label{eq:composition}
  \textnormal{.}
\end{align}

\subsection{All-$1$s Driving}
\label{subsec:all0s}

\begin{figure}[ht]
\centering
\includegraphics{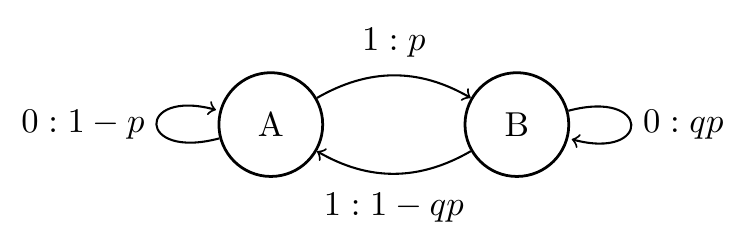}
\caption{Output HMC given by the AS4C transducer acting on the all-$1$s
	input tape. The ratchet in this case scrambles an informationless
	input, thereby introducing the capacity to do work.
	}
\label{fig:as4c_ao_ohmm}
\end{figure}

To simplify determining $h'_\mu$, we drive the AS4C transducer with the
\emph{all-ones process}: an input tape of all $1$s, exhibiting no randomness
whatsoever. Note that generically the output HMC of a memoryless ratchet driven
by a memoryless input process results in a highly nonunifilar output HMC
\footnote{Nonunifilarity refers to the chosen symbol allowing transitions to
multiple next states.}, for which determining the entropy rate is very
challenging \cite{jurgensFunctionalThermodynamicsMaxwellian2020}. However, for
all-$1$s driving, the AS4C produces the unifilar output HMC shown in
Fig.~\ref{fig:as4c_ao_ohmm}.

\begin{figure}[ht]
\centering
\includegraphics{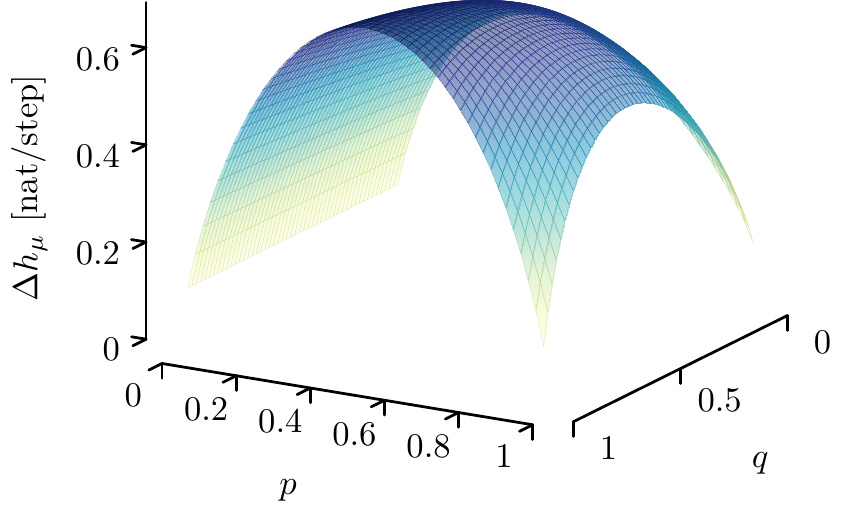}
\caption{Change in Shannon entropy rate $\Delta h_\mu$ of the information tape,
	as generated by the AS4C ratchet driven by the all-$1$s process. Since the
	input process is entirely ordered with $h_\mu = 0$, this represents the ESS
	IPSL's maximum upper bound $-\beta \dot{\expval{Q}}_\textnormal{min}$. And
	this, in turn, precludes eraser functionality. That is, one cannot erase
	information that was never there.
	}
\label{fig:AS4C_hmu_AO}
\end{figure}

Since this HMC is unifilar---an internal state and an output symbol completely
determine the next internal state---and since its two states make
probabilistically distinct future predictions, it is a finite-state \eM\ of
computational mechanics \cite{crutchfieldOrderChaos2012}.  That the output
tape's process can be described this way enables direct calculation of the
output entropy rate \cite{crutchfieldRegularitiesUnseenRandomness2003}. Letting
$i$ and $j$ index the output HMC's internal states, $\bm{\pi}$ be its
stationary distribution, and $y$ an output symbol, one has:
\begin{align}
  h'_\mu &= -\sum_{y, i, j} \pi\pqty{i} V^\pqty{y}_{ij} \log V^\pqty{y}_{ij}
  \label{eq:hmu_uHMM}
  \textnormal{.}
\end{align}
Figure~\ref{fig:AS4C_hmu_AO} plots this over the parameter space.

Setting the input process to have zero randomness also sets $h_\mu = 0$ for it:
$\Delta h_\mu = h'_\mu$, all intrinsic randomness in the output tape is induced
by the ratchet and, therefore, is available as a thermodynamic resource for heat
extraction. For the case of totally ordered input,
Eq.~\eqref{eq:asymptotic_IPSL_simplified} reads:
\begin{align}
  -\beta \dot{\expval{Q}} \leq h'_\mu - \dot{\expval{\gQhk}}
  \textnormal{.}
\end{align}
Summarizing the requirements for net heat extraction: the ratchet must, at
minimum, induce randomness in the output tape faster than it dissipates entropy
to maintain its NESS.

Since with this particular driving $\Delta h_\mu > 0$ for all parameter
combinations, the information eraser functionality of
Refs.~\cite{boydIdentifyingFunctionalThermodynamics2016,
jurgensFunctionalThermodynamicsMaxwellian2020} is precluded. Instead, we have
either a heat engine ($\expval{Q} < 0$) or a dud ($\expval{Q} \geq 0$). Most
importantly, the presence of $\dot{\expval{\gQhk}}$ here restricts the regions
of parameter space where the ratchet \emph{can} function as a heat engine. Or,
alternatively, for some $p$ and $q$ the housekeeping costs are higher than the
ratchet's ability to compensate by scrambling the information tape. This forces
the previous ``potential engine'' regions into dud regions.

\begin{figure}[ht]
\centering
\includegraphics{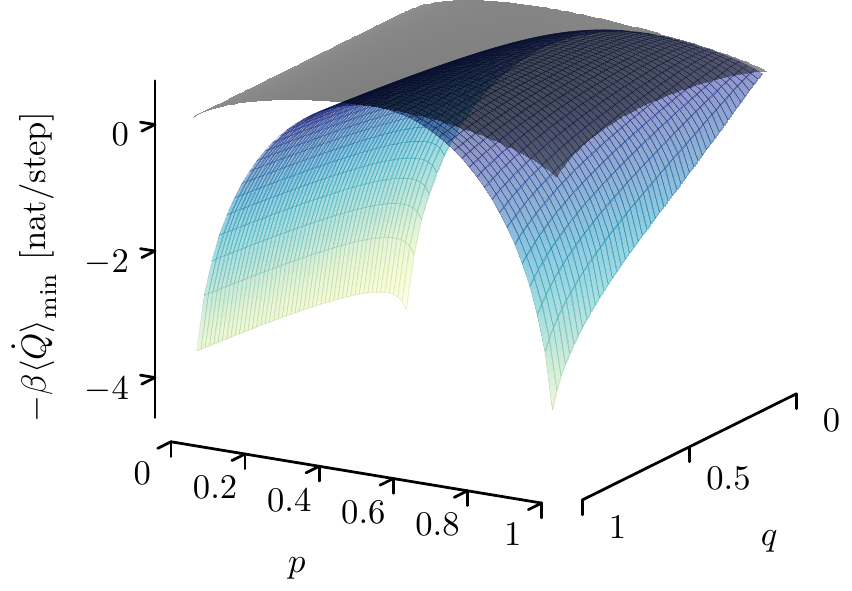}
\caption{NESS-tightened upper bound on heat extraction $-\beta
	\dot{\expval{Q}}_\textnormal{min} = \Delta h_\mu - \dot{\expval{\gQhk}}$ for
	the AS4C driven by the all-$1$s process. To the extent that this differs
	from Fig.~\ref{fig:AS4C_hmu_AO}, superimposed here in gray, it represents a
	change of maximum possible net heat extraction.
	}
\label{fig:AS4C_Wmax_AO}
\end{figure}

\begin{figure}[ht]
\centering
\includegraphics{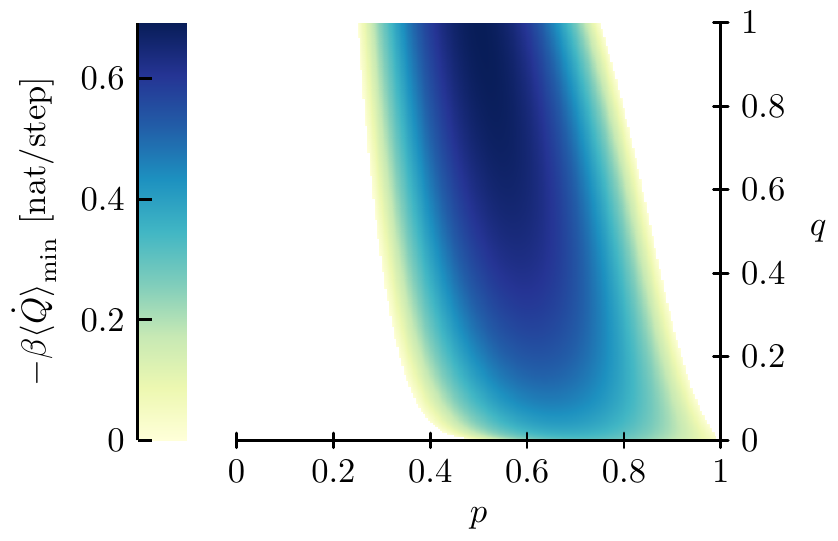}
\caption{Parameter space regions that permit the all-$1$s-driven AS4C ratchet to
	function as a heat engine. Notably, this includes only a band centered
	around detailed-balanced dynamics. The remainder of the uncolored parameter
	space forces $-\beta \dot{\expval{Q}}_\textnormal{min} < 0$, where the ratchet dissipates heat on average.}
\label{fig:AS4C_Wmax_AO_2D}
\end{figure}

This is indeed the case, as Fig.~\ref{fig:AS4C_Wmax_AO} shows. In fact, only a
small part of engine functionality remains within bounds. Figure
\ref{fig:AS4C_Wmax_AO_2D} shows this directly, where only ``potential engine''
regions of parameter space are colored. Since an entirely ordered input drives
the ratchet, without accounting for the NESS correction one would expect all
parameter space to allow potential heat extraction. In this way, explicitly
accounting for a system's NESS nature enables qualitative (and quantitative)
correction to its allowed behaviors.

\section{Conclusion}
\label{sec:conclusion}

We began by deriving, under very general circumstances, an IPFL that connects
ensemble-averaged thermodynamic behavior to a system's information processing
via a strict equality. We showed that this equality is, equivalently, a
decomposition of stochastic thermodynamics' average total entropy production. To
get there, we placed very few restrictions on the underlying system's dynamics,
considering transitions between nonequilibrium dynamical states.

From this First Law, we then applied integral fluctuation theorems to take the
equalities to inequalities, reproducing and then tightening established bounds
on average heat extraction. By splitting the system into ratchet and tape
subspaces and considering both finite and infinite-time cases, we similarly
reproduced and then tightened previous IPSLs of autonomous Maxwellian ratchets
\cite{boydIdentifyingFunctionalThermodynamics2016,boydLeveragingEnvironmentalCorrelations2017,
jurgensFunctionalThermodynamicsMaxwellian2020,
heInformationProcessingSecond2022} to explicate the effects of nonequilibrium
dynamical states. Finally, we illustrated these results with an example
ratchet-tape system---the AS4C, driven by the ordered all-$1$s process. This
demonstrated that, even under extreme simplification, the presence of NESSs
introduced qualitative corrections to a ratchet's allowed behavior. In short,
the presence of housekeeping entropy costs, induced by NESSs, directly
counteracts a ratchet's ability to leverage information creation to extract
energy from a heat bath.

Much room for further development remains, particularly in light of the role of
fluctuation theorems in deriving these IPSLs. While our derivation concerned
full ensemble averages, recent development of trajectory-class fluctuation
theorems
\cite{semaanHomeostaticAdaptiveEnergetics2022,wimsattTrajectoryClassFluctuation2022}
highlight opportunities to derive trajectory class IPSLs that are more amenable
to experimental verification via their freedom from rare-event statistical
errors \cite{jarzynskiRareEventsConvergence2006}.

That odd-parity variables allow for meaningful decomposition of the housekeeping
heat suggests further explication of their effects on the derived IPFLs and
IPSLs, including bounds on asymptotic work extraction. Indeed, recent results in
stochastic thermodynamics show that where a known, constrained splitting of the
joint state space is available, it may be used to tighten the corresponding
Second Laws \cite{wolpertStrengthenedSecondLaw2022}. Additionally, considering
that infinite-state ratchets revealed new contributions to the underlying IPSL,
their convergence or divergence in general cases warrants detailed analytical
investigation.

Finally, the fact that the NESS setting does not straightforwardly account for
work extraction itself warrants further study. Indeed, even without the formal
challenges surrounding work calculation, the presence of three thermodynamic
reservoirs implies additional net fluxes. While the result remains that
information-bearing degrees of freedom can---in principle---provide a
thermodynamic resource, the presence of additional dissipation from NESSs
suggests that care must be taken when determining information engine
efficiency: two ratchets that produce the same asymptotic work may nevertheless
produce (potentially very) different heats.

Taken together, these results demonstrate that combining familiar
tools---average change in steady-state surprisal and a single integral
fluctuation theorem---simplifies and generalizes deriving IPSLs. In turn,
these bound the extent to which systems can leverage information-bearing
degrees of freedom to support thermodynamic functionality. Furthermore, we
showed explicitly how such inequalities arise from underlying equalities. This
appeared in much the same way as stochastic thermodynamics' fluctuation
theorems simplify to the original statement of the Second Law
\cite{jarzynskiEquilibriumFreeenergyDifferences1997,
jarzynskiNonequilibriumEqualityFree1997}.

\section*{Acknowledgments}
\label{sec:acknowledgements}

The authors thank Gregory Wimsatt, Alec Boyd, Paul Riechers, Kyle Ray, Samuel
Loomis, Alex Jurgens, Adam Rupe, Jacob Hastings, and Nicolas Brodu for
illuminating discussions, as well as the Telluride Science Research Center for
its hospitality during visits and the participants of the Information Engines
workshop there for their valuable feedback. J.P.C. acknowledges the kind
hospitality of the Santa Fe Institute, Institute for Advanced Study at the
University of Amsterdam, and California Institute of Technology. This material
is based on work supported by, or in part by, Foundational Questions Institute
Grant No. FXQi-RFP-IPW-1902 and the U.S. Army Research Laboratory and U.S. Army
Research Office under Grants No. W911NF-18-1-0028 and W911NF-21-1-0048.

\appendix
\section{Trajectory versus State Averaging}
\label{apdx:avg_proof}

The main result relies on the equivalence between $\expval{\Delta \phi_\param}$
and $\Delta \braket{\mu}{\phi_\param} = \braket{\mu_N}{\phi_{\param_N}} -
\braket{\mu_0}{\phi_{\param_0}}$. The former refers to $\Delta \phi_\param$'s
average over an ensemble $\{ z_{0:N} \}$ of repeated trajectories and, thus,
means $\expval{\Delta \phi_\param} = \expval{\gWex} - \expval{\gQex}$. The
latter refers to two specific state averages of $\phi_\param$---namely, those at
the trajectory's endpoints. And, it is equal to $\Delta \Hent{Z} + \Delta
\DKL{Z}{\Lambda}$, via the arguments in Eq.~\eqref{eq:derivation_HplusD}. We
establish the equivalence between the path and state averages of $\Delta
\phi_\param$ here.

The \textit{trajectory average} of a path-dependent functional
$g : \mathcal{Z}^N \to \mathbb{R}$, denoted $\expval{g}$, is:
\begin{align}
  \expval{g} \defn \int g\pqty{z_{0:N}}\, \pr{z_{0:N}}
    \pqty{\prod_{i=0}^N \dd{z_i}}  
  \textnormal{.}
\end{align}
We will extend this definition to functionals from $\mathcal{Z}^n \to
\mathbb{R}$ for integer $n$, where $1 \leq n < N$, by simply placing the
functional in the integral above while keeping the average over the full
trajectory space $\mathcal{Z}^N$.

The \textit{state average} of a function $f : \mathcal{Z} \to \mathbb{R}$, denoted
$\braket{\mu_i}{f}$, is:
\begin{align}
  \braket{\mu_i}{f} \defn \int f\pqty{z_i}\, \pr{z_i}
    \dd{z_i}  
  \textnormal{.}
\end{align}

\textbf{Claim.} For any $f\pqty{z_n}$ that depends only on one point $z_n$, $0
\leq n \leq N$ in the path, the path and state averages are equal: $\expval{f} =
\braket{\mu_n}{f}$.

\begin{proof}
\newlength{\myindent} \settowidth{\myindent}{(ii) } We explicitly evaluate the trajectory average. Consider two cases: (i) $n = N$,
and (ii) $0 \leq n < N$.
  \begin{enumerate}[leftmargin=\myindent, label=(\roman*)]
    \item First, split the path probability into two pieces: $\pr{z_{0:N}}
    = \pr{z_{0:N-1}}\cpr{z_N}{z_{0:N-1}}$. Now, evaluate the integrals for
    $\dd{z_0}$ through $\dd{z_{N-1}}$:
      \begin{align*}
        &\int \pr{z_{0:N-1}}\cpr{z_N}{z_{0:N-1}}
          \pqty{\prod_{i=0}^{N-1} \dd{z_i}} \\
          &\qquad = \pr{z_{N}}
        \textnormal{,}
      \end{align*}
by the law of total probability. The remainder is the $\dd{z_N}$ integral:
      \begin{align*}
        \int \pr{z_N} f\pqty{z_N} \dd{z_N} = \braket{\mu_N}{f}
        \textnormal{,}
      \end{align*}
by definition.

    \item Again split the probability, but now as $\pr{z_{0:N}} =
    \pr{z_{0:n}} \cpr{z_{n+1:N}}{z_{0:n}}$. Evaluate the integrals
    for $\dd{z_{n+1}}$ through $\dd{z_N}$:
      \begin{align*}
        \int \cpr{z_{n+1:N}}{z_{0:n}}
          \pqty{\prod_{i=n+1}^{N} \dd{\xx_i}}
          = 1
      \end{align*}
by probability conservation. What remains is exactly case (i).
  \end{enumerate}
\end{proof}

This assumes a truly finite stochastic process, such that no conditioning
before $z_0$ or after $z_N$ is possible or relevant. However, the result is
robust in the limit of a bi-infinite stochastic process. Evaluating the future
integral in (ii) still yields $1$ in the $N \to \infty$ limit. And, then, the
past integral in (i) still gives $\pr{z_n}$, even as the lower bound extends to
$-\infty$.

Furthermore, we did not require Markovity, ergodicity, or stationarity for
the underlying stochastic process. The result, then, appears quite general. This
is not too surprising: a point function's average over paths should not depend
on the path. Yet the link between path-independent ($\Delta \phi_\param$) and
path-dependent ($\gWex$ and $\gQex$) quantities provided by the nonaveraged
First Law renders it particularly useful.


\end{document}